\documentclass{WileyMSP-template}

\usepackage{graphicx}%
\usepackage{multirow}%
\usepackage{amsmath,amssymb,amsfonts}%
\usepackage{amsthm}%
\usepackage{mathrsfs}%
\usepackage[title]{appendix}%
\usepackage{xcolor}%
\usepackage{textcomp}%
\usepackage{manyfoot}%
\usepackage{booktabs}%
\usepackage{algorithm}%
\usepackage{algorithmicx}%
\usepackage{algpseudocode}%
\usepackage{listings}%

\usepackage{tikz-cd}
\usepackage{parskip}

\usepackage{geometry} 
\newtheorem{Definition}{Definition}

\begin{document}
	
\pagestyle{fancy}
\rhead{\includegraphics[width=2.5cm]{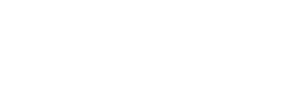}}
	
\title{Measurements on the separated subsystems of an entangled state}
\maketitle

\author{Gregory D. Scholes}

\begin{affiliations}
	G. D. Scholes\\
	Department of Chemistry, Princeton University, Washington Rd. Princeton, New Jersey 08544, U.S.A.\\
	Email Address: gscholes@princeton.edu
\end{affiliations}


\begin{abstract}
	
	The entangled states of composite quantum systems are well studied. The particle-like nature of these systems also means that, while entangled, they can be physically separated and measurements performed on the separated subsystems. Measurements on the separated subsystems A and B should pertain to vectors in the local Hilbert space of each subsystem, but to date it has not been clear how to elucidate the relevant states of the separated subsystems because it is not obvious how to resolve them from states given in the tensor product basis, except for the separable states. Here it is shown that the projections of any general entangled state that are detected by measurements on the separated subsystems can be obtained considering the corresponding (cosets of) states in the free vector space from which the tensor product space is defined. The result eliminates the need to invoke random collapse, and from this perspective nonlocality arises because of the way measurements on each separated subsystem projects possible measurement outcomes.
	
\end{abstract}

\date{\today}

\newpage

\section{Introduction}

Here we examine measurements on separated subsystems associated with an entangled state. While it is easy to see how measurements on the separated subsystems relate to a simple tensor state of the composite system\cite{vonNeumann}, it is not obvious how to relate those measurements to a general entangled state. The main idea proposed here is that physical separation of the subsystems enables the state of a composite system to `project' (in the abstract sense) into the Hilbert spaces local to the subsystems. Based on the definition of the tensor product, it is shown how to obtain such a `projection'. This gives insight into how  separation of the subsystems allows an entangled state to be probed by the results of classical measurement read-outs. In addition, insight into the origin of nonlocal correlations in obtained.

When constructing the states of composite quantum systems\cite{Peres} from the state spaces $U$ and $V$ of their subsystems A and B, we require: (i) that the resulting state space is linear with respect to the action of maps, and (ii) linear operators on the subsystems, $T_A$ acting on $U$ and $T_B$ acting on $V$, extend to a linear operator acting on the composite state. That is accomplished by defining states in the tensor product space $\mathcal{T} = U \otimes V$. 

The space $U \otimes V$ contains not only simple tensors (e.g. $u \otimes v$), but also linear combinations of tensors. A subset of the latter are the maximally entangled states. A prototypical example is the singlet state
\begin{equation}
	\Psi_- = \frac{1}{\sqrt{2}} \Big( |0\rangle_A |1\rangle_B - |1\rangle_A |0\rangle_B  \Big),
\end{equation}
where $|0\rangle_A$ and $|1\rangle_A$ are basis vectors for subsystem A in $\mathcal{H}_A$, etc.

The properties of $U \otimes V$ and the states in this space, the states of composite quantum systems, are well studied. However, quantum mechanics is a theory for systems like electrons, photons, and so on, whose states can be entangled and lie in $U \otimes V$. But, the particle-like nature of these systems also means that, while entangled, they can be physically separated and measurements performed on the separated subsystems. Measurements on the separated subsystems A and B should pertain to vectors in $\mathcal{H}_A$ and $\mathcal{H}_B$, but to date it has not been clear how to elucidate the relevant states (of the separated subsystems) because it is not obvious how to resolve them from states in the tensor product basis. The present paper explores this topic.

Here we derive the relationship between state vectors of the entangled system (linear combinations of tensors) in the composite Hilbert space $\mathcal{H}_A \otimes \mathcal{H}_B$, and state vectors of the separated subsystems in $\mathcal{H}_A$ and $\mathcal{H}_B$. Von Neumann recognized that this is the approach needed\cite{vonNeumann}. As he indicates in Chapter VI.2 of \cite{vonNeumann}, such measurement outcomes correspond to the expectation values in the Hilbert spaces of the separated subsystems, $\mathcal{H}_A$ or $\mathcal{H}_B$. Also see Ref. \cite{HughsBook} Chapter 5.7. Von Neumann showed the expected  correspondence between measurements on subsystem A in $\mathcal{H}_A \otimes \mathcal{H}_B$ compared to in $\mathcal{H}_A$ for the case of separable composite states. That is, he used the fact that eigenvalues $\lambda_i$ for some operator $T_A$ and vectors $u_i^A \in  \mathcal{H}_A$:
\begin{equation*}
	T_A u_i^A = \lambda_i u_i^A,
\end{equation*}
are also eigenvalues for the related operator on states in $\mathcal{H}_A \otimes \mathcal{H}_B$:
\begin{equation*}
	(T_A \otimes \mathbb{I}_B) (u_i^A \otimes v_j^B) = \lambda_i (u_i^A \otimes v_j^B),
\end{equation*}
where $\mathbb{I}_B$ is the identity operator on subsystem B. That is, an operator $T_A$ acting on $u_i^A \in \mathcal{H_A}$ gives the same eigenvalues $\lambda_i$ as $T_A \otimes \mathbb{I}_B$ acting on $u_i^A \otimes v_j^B$.

On the other hand, how measurements happen on entangled states $\Psi$ is less clear. First, $\Psi$ is not an eigenvector of $T_A \otimes \mathbb{I}_B$. Second, the operator only gives one outcome for a measurement, whereas we can obtain two measurement outcomes if we probe each subsystem individually. Third, when we approach this problem, it appears that we need to invoke some kind of random collapse model to obtain definite measurement outcomes for the subsystems\cite{OmnesBook}. In the present work it is shown how to project $\Psi \in \mathcal{H}_A \otimes \mathcal{H}_B$ into the subsystem Hilbert spaces $\mathcal{H}_A$ and $\mathcal{H}_B$ to obtain measurement outcomes for the separated subsystems. The results eliminate the need to invoke random collapse, and gives the insight that nonlocality arises because of the way measurements on each separated subsystem projects possible measurement outcomes.

\section{Technical background}

For the main results in this paper, we will need to work with, not only the vectors in the tensor product space $\mathcal{T}(U_A \otimes V_B)$, but also in the covering free vector space $\mathcal{F}(U_A \times V_B)$. We start by giving background on free vector spaces and develop some intuition for the relationship between free vector spaces and their quotient spaces, the tensor product spaces. The main result of the paper is worth keeping in mind: In order to use the canonical projection to obtain states in the Hilbert spaces corresponding to the separated subsystems, we need an explicit Cartesian product basis \emph{and} we need arbitrary finite linear combinations (to accommodate entangled states). Hence we need to consider representatives of a state in the relevant free vector space that reduces to our tensor product space by action of a quotient. 

The following discussion is restricted to finite-dimensional vector spaces. The vector spaces considered can be over any field, but we use the explicit example where the field is the complex numbers ($\mathbb{C}$). We begin with some technical details. Throughout, $\alpha, \beta$ refer to scalars in our field, $\mathbb{C}$. Using the convention in mathematics, we will often specify a state in a vector space without using Dirac notation. That is, we consider the notation $\psi$ equivalent to $| \psi \rangle$. 

\begin{Definition}{(Cartesian product of sets)}
	Let $S_1, \dots, S_n$ be any sets. The Cartesian product of sets, denoted $S_1 \times \dots \times S_n$, is the set of all ordered $n-$tuples $a_1, \dots, a_n$, where $a_i \in S_i$ for $i = 1, \dots n$.
\end{Definition}

\begin{Definition}{(Free vector space)}
	A free vector space $\mathcal{F}(\mathcal{S})$ on a set $\mathcal{S}$ over a field $\mathbb{K}$ is a vector space comprising all finite formal linear combinations of elements in $\mathcal{S}$, where the elements of $\mathcal{S} = \{ e_0, e_1, \dots \}$ form a basis.
\end{Definition}

The notion of formal linear combinations means that we have linear combinations of the form $\alpha_0 e_0 + \alpha_1 e_1 + \dots$ ($\alpha_i \in \mathbb{K}$), but no relations among the symbols (i.e. $e_0, e_1, \dots$) are assumed.  By defining the free vector space over $\mathcal{S}$, construction of formal linear combinations of symbols, the elements of $\mathcal{S}$, become analogous to linear combination of basis elements, say $|0\rangle, |1\rangle, \dots$ of a vector space such as a Hilbert space $\mathcal{H}$. Those basis elements are vectors that satisfy certain relations so that $\psi = |0\rangle + |1\rangle$ is equivalent to $\psi' = 3|0\rangle - 2|0\rangle + |1\rangle$ because we can compute sums of a vector. Similarly, in a free vector space $\mathcal{F}(\mathcal{S})$ the elements $e_0 + e_1$ and $3e_0 -2e_0 + e_1$ are equivalent.  

To clarify the nature of this map $\mathcal{F} \rightarrow \mathcal{H}$, we define a free vector space $\mathcal{F}(U)$ such that its elements map bijectively to the Hilbert space vectors of a quantum subsystem $\psi = \alpha|0\rangle + \beta|1\rangle$. That is, we take our basis to be $\mathbb{Z}_2 = \{0, 1\}$, written $\{e_0, e_1\}$, and define the map $e_0 \rightarrow |0\rangle$ and $e_1 \rightarrow |1\rangle$. Define $\mathcal{F}(U)$ over $\mathbb{C}$ as all finite $\mathbb{C}$-linear combinations of the formal basis functions $e_0$ and $e_1$ so that $\mathcal{F}(U) \cong \mathbb{C}^2$ with elements $\alpha e_0 + \beta e_1$. Choose the identification map $T: \mathcal{F}(U) \rightarrow \mathcal{H}$ such that $T(e_0) = |0\rangle$ and $T(e_1) = |1\rangle$, where $|0\rangle$ and $|1\rangle$ form the basis of $\mathcal{H}$. Then extend linearly, so that $T(\alpha e_0 + \beta e_1) =  \alpha|0\rangle + \beta|1\rangle$. 

Now define $\mathcal{F}(U \times V)$ as the free vector space over the field $\mathbb{C}$ on the set $U \times V$ (i.e. Cartesian product). It has elements that are finite $\mathbb{C}$-linear combinations of basis symbols and satisfies the universal property for maps from $U \times V$ into $\mathbb{C}$-vector spaces. 

The tensor product\cite{TensorSpaces, WarnerBook, Roman} is defined from the free vector space as follows. Let $\mathcal{F}(U \times V)$ be the free vector space over the field $\mathbb{C}$ with basis $U \times V$. In the following we implicitly associate elements of $\mathcal{F}(U \times V)$ to corresponding vectors. Let $\mathcal{R}$ be the subspace of $\mathcal{F}(U \times V)$ generated by all vectors of the form
\begin{eqnarray}
	r(u, w) + s(v, w) - (ru + sv, w) \\
	r(u, v) + s(u, w) - (u, rv + sw) \nonumber 
\end{eqnarray}
with $s, r \in \mathbb{C}$. These vectors are what we must identify as the zero vector in order to ensure bilinearity of a map on a vector in $U \times V$ equates to a linear map of a corresponding vector in $U \otimes V$, which is the key requirement for defining the tensor product. We now define the tensor product by the quotient
\begin{equation}
	\mathcal{T} = U \otimes V := \mathcal{F}(U \times V)/\mathcal{R} .
\end{equation}

The tensor product space  $\mathcal{T}(U \otimes V)$ is related to $\mathcal{F}(U \times V)$ by the universal mapping property\cite{Roman, TensorSpaces}:
\[
\begin{tikzcd}[column sep=large, row sep=large]
	U \times V \ar[rd, "f "'] \ar[r, "t"] & U \otimes V  \ar[d, description, "\tau"'] \\
	& W
\end{tikzcd}
\]
Here the map $\tau$ is a linear map, whereas $f$ and $t$ are bilinear maps. That is, by forming the tensor product space we linearize the bilinear maps on $U \times V$. A simple example to illustrate this is a point $(x,y)$ on the Cartesian plane $\mathbb{R} \times \mathbb{R}$. If we form the tensor product space $\mathbb{R} \otimes \mathbb{R}$, which is isomorphic to $\mathbb{R}$, then the bilinear maps on the plane become linear maps on the line. Then the points $(-12, 4)$, $(4,-12)$, $(-3, 16)$, and so on, in $\mathbb{R} \times \mathbb{R}$ all map to one point, e.g. $48$, in $\mathbb{R} \otimes \mathbb{R} \cong \mathbb{R}$. Conversely, specifying a point on the line (e.g. 48) does not contain enough information to indicate uniquely a single corresponding point in $\mathbb{R} \times \mathbb{R}$. Instead it denotes a set of points, its equivalence class with respect to the quotient map. 

The free vector space covers the tensor product space in an algebraic sense because $\mathcal{F}(U \times V)$ maps surjectively onto $U \otimes V$ via the quotient map. However, we have to be careful. In $U \otimes V$ we can distinguish entangled states from linear combinations that can be expressed as a single tensor product by using bilinearity and the distributive law. Whereas, we cannot make the same distinction between points in the free vector space $\mathcal{F}$, because they are denoted only by a formal sum of independent points, as emphasized above. So in the following, we are going to consider the way tensors in $U \otimes V$ relate to equivalence classes of points in $\mathcal{F}$. That way we retain the structure of the tensor product space, but we can identify representatives in $\mathcal{F}$. This identification is not a map. Instead, a way to think about it is that the wavefunction $\Psi$ of an entangled state hides these representatives because it is defined in a quotient space. Whereas, the representatives are exposed when measurements are made on the separated subsystems, as will be shown below. This way of thinking about $\Psi$ enables, in the next section of the paper, understanding of why the measurement outcomes are definite, and why measurement outcomes are correlated among the subsystems.

A potential point of confusion is that it is well advertised that tensor product spaces are often much larger than corresponding Cartesian product spaces. That is the case because\cite{KadRing1}, if $U$ and $V$ have dimension $m$ and $n$ respectively (each being $\ge 2$), then $\dim U \otimes V = mn$ and $\dim U \times V = m+n$. Then we can gain insight by counting the number of possible states generated by all the possible superpostions with coefficients restricted equal 1 or 0 (i.e. the coefficients are elements of $\mathbb{F}_2$). Then we can compare the size of the state spaces to find that $|U \otimes V| = 2^{mn}$ and $|U \times V| = 2^{m+n}$. The calculation is simply the sum of binomial coefficients. Now, the free vector space $\mathcal{F}(U \times V)$ has one basis element for each $|U \times V|$. Therefore, in our example, $|\mathcal{F}(U \times V)| = 2^{2^{m+n}}$, which is vastly larger than is vastly larger than $|U \otimes V|$.  That is, the quotient map identifies sets of many distinct formal linear combinations that map to the same tensor in $U \otimes V$. Finally, note that the number of basis elements in $\mathcal{F}(U \times V)$ is the same as in $U \times V$ and $U \otimes V$. That is, we can send every $(u,v)$ to a unique simple tensor $u \otimes v$.

\section{Measurements on the states of the composite system}

To begin we give the necessary approach and show that measurement outcomes for operators acting on the states of a composite system are the same whether we describe them as measurements on vectors in $\mathcal{T}$ or $\mathcal{F}$.  Consider the entangled states of composite quantum systems, where the subsystems are labelled A and B. We have $\mathcal{F}(U_A)$, $U_A = \{ e_0, e_1 \}$ that is related to the Hilbert space of subsystem A by an identification map $T(\alpha e_0 + \beta e_1) = \alpha |0\rangle + \beta |1\rangle \in \mathcal{H}_A$, and similarly for subsystems B. 

The identification map for a point in $\mathcal{F}(U_A \times V_B)$ is the corresponding bilinear identification map (see Sec. 9 of Ref. \cite{TensorSpaces}) $t\big((u,v)\big) = |u\rangle \otimes |v\rangle$, which is extended linearly to accommodate formal linear combinations in $\mathcal{F}(U_A \times V_B)$ that map to linear combinations of pure tensors in $\mathcal{T}(U_A \otimes V_B)$. The map on any representative of the equivalence class $[(u,v)]$ in $\mathcal{F}$ gives the same single vector in $\mathcal{T}$. For example, $t\big((-u,v)\big) = -|u\rangle \otimes |v\rangle = t\big((u,-v)\big)$.

For an explanatory example, consider $U = V = \mathbb{R}^2$, so that $U \times V$ is 4-dimensional Cartesian space. Each $U, V$ has a basis $\{e_0, e_1\}$. An element in $\mathcal{F}(U \times V)$ is any formal sum, for instance $\psi = (3e_0, 2e_0) + 4(e_1, e_0) + (e_1, -2e_1)$. The image of $\psi$ in $U \otimes V$ is $t(\psi) = 6e_0 \otimes e_0 + 4 e_1 \otimes e_0 - 2 e_1 \otimes e_1$.

Now we can formally compare the action of a linear operator $T_A \otimes T_B$ on states $u \otimes v + x \otimes y$ in $\mathcal{T}$ with the action on the corresponding formal linear combination $(u,v) + (x,y)$ in $\mathcal{F}$. Of course, the results must be the same, as a consequence of the notion of universality with respect to a bilinear map\cite{Roman, TensorSpaces}. Any bilinear map $f: U \times V \rightarrow W$ can be factored through $t: U \times V \rightarrow U \otimes V$, as indicated by the composition of commuting maps given above.

We know that 
\begin{equation*}
	(T_A \otimes T_B)(u \otimes v + x \otimes y) = T_A(u) \otimes T_B(v) + T_A(x) \otimes T_B(y) ,
\end{equation*}
and that an observable of subsystem A in the composite system, therefore, is probabilistically
\begin{align}
	E_A &= \frac{1}{2} \langle u \otimes v|T_A \otimes \mathbb{I}_B |u \otimes v  \rangle + \frac{1}{2} \langle x \otimes y|T_A \otimes \mathbb{I}_B |x \otimes y  \rangle \\
	&= \frac{1}{2} \langle u |T_A |u \rangle + \frac{1}{2} \langle x |T_A |x  \rangle .
\end{align}
Here we considered a case where cross-terms can be ignored to simplify the example.

The same result is evident for the operator $T_A \times T_B$ acting on $(u,v) + (x,y)$ in $\mathcal{F}$. To see this, note that the map acts to give $\big(T_A(u),T_B(v) \big) + \big(T_A(x),T_B(y) \big)$. Use the identification map defined above to obtain corresponding vectors in the Hilbert space, which is the tensor product space. Thus we obtain the same result.

Therefore, the measurement outcomes are the same whether we work with states in $\mathcal{T}$ or $\mathcal{F}$. This is not a surprise, but is helpful to establish before proceeding to the next section. We also point out that we can take \emph{any} representative of an equivalence class in $\mathcal{F}$ and obtain precisely the same expectation values---in contrast to the results to be reported in the following section.  Finally, note that a measurement on a state of the composite system gives one measurement outcome.

\section{Measurements on separated subsystems}

Entangled states are characteristic of quantum systems. However, while the \emph{states} of a quantum system can be entangled, the physical subsystems that comprise the composite system can be physically separated. It is this separation of the subsystems, so that they remain entangled but do not interact with each other, that lies behind interesting correlations in the measurements of the subsystems\cite{Mermin1985, Peres, Laloe, Ballentine, OmnesBook, BellBook, CS1978, GHSZ1990, ReidEPR, hidden10, Selleri, Brunner, Aspect1982, Aspect2015, Kupczynski2006}. One of the underlying reasons for the interest and importance of measurements on the separated subsystems is that, while a measurement on the composite state gives one observable---one measurement outcome---we obtain \emph{two} observables by measurements on the separated subsystems. 

In isolation, the states of the subsystems A and B lie in the Hilbert spaces $\mathcal{H}_A$ and $\mathcal{H}_B$, respectively. Hence, separating the subsystems associated with an entangled state must somehow enable measurements on a `projection' of $\mathcal{H}_A \otimes \mathcal{H}_B$ into $\mathcal{H}_A$ and $\mathcal{H}_B$. Note that this is an abstract notion of projection that does not imply a map using a projection operator. In fact, there is no such map from the tensor product space because $\mathcal{F}(\mathcal{H}_A \times \mathcal{H}_B) \rightarrow \mathcal{H}_A \otimes \mathcal{H}_B$ is a surjective map via the quotient. Then, how can we find a suitable well-defined `projection' from an entangled state into $\mathcal{H}_A$ and $\mathcal{H}_B$? 

The desired `projection' cannot be accomplished for general states in the tensor product space $U \otimes V$, where the basis for each subsystem has been converted to a new basis for the composite system. Instead, we need to explain measurements on the separated subsystems in terms of expectation values of the subsystem states in their respective Hilbert spaces $\mathcal{H}_A$ and $\mathcal{H}_B$;  the vectors in these spaces are precisely those associated with separated subsystems.  We can capture this concept in the following way:

(Measurements of separated subsystems) Let $\Psi$ be a bipartite maximally entangled state specified in the tensor product space $\mathcal{T}(\mathcal{H}_A \otimes \mathcal{H}_B)$, where $\mathcal{H}_A$ is the Hilbert space of states for subsystem A and $\mathcal{H}_B$ is the Hilbert space of states for subsystem B. Measurement outcomes on the subsystems A and B when separated from $\Psi$ are the observables associated with projections of representatives of the equivalence class $[\Psi] \in \mathcal{F}(U_A \times V_B)$.

The hypothesis is justified by first recalling that the vector $\Psi$ is specified in a quotient space, 
\begin{equation*}
	\mathcal{T}(\mathcal{H}_A \otimes \mathcal{H}_B) = \mathcal{F}(U_A \times V_B)/\mathcal{R},
\end{equation*}
where $\mathcal{R}$ is the subspace given in Eq. 2. 

Notice that the elements of $\mathcal{T} = U \otimes V$ are expressed in terms of elements of $\mathcal{F}$ by
\begin{equation}
	\big( \sum r_i (u_i, v_i) \big) + \mathcal{R} = \sum r_i \big[ (u_i, v_i) + \mathcal{R} \big]
\end{equation}
where $[\dots]$ means equivalence class. Since $r(u, v) - (ru, v) \in \mathcal{R}$ and $r(u, v) - (u, rv) \in \mathcal{R}$ we can always absorb the scalar in either coordinate, that is
\begin{eqnarray}
	r \big[ (u,v) + \mathcal{R} \big] &= r(u \otimes v) \nonumber \\
	&= (ru, v) + \mathcal{R} \\
	&= (u, rv) + \mathcal{R}.
\end{eqnarray}
These equivalences are irrelevant for measurements on the composite system, but become important when we consider separated subsystems. 

As such, $\Psi$ denotes an equivalence class of representative vectors $[\Psi]$ in the (algebraic) covering space $\mathcal{F}(U_A \times V_B)$. The formal linear combinations of tuples in $\mathcal{F}(U_A \times V_B)$ have a bijective correspondence to the obvious Cartesian product basis implied by $\mathcal{H}_A \times \mathcal{H}_B$. Therefore, observables associated with measurements on the separated subsystems A and B are expectation values of vectors respectively associated with $\mathcal{H}_A$ and $\mathcal{H}_B$ by canonical projection from $\mathcal{F}(U_A \times V_B)$ defined as follows. 

The canonical projection from an element of $\mathcal{F}(U \times V)$ to an element of $\mathcal{F}(U)$, $\pi_U : \mathcal{F}(U \times V) \rightarrow \mathcal{F}(U)$ is the linear extension of $(u,v) \rightarrow u$, such that
\begin{equation}
	\pi_U\Big( (u,v) + (x,y) \Big) = u + x \quad \in \mathcal{F}(U).
\end{equation}

Now, to obtain the corresponding vector in the Hilbert space of the subsystem, for finite spaces, we bijectively assign the basis vectors $v_u$ of $U$ to corresponding formal basis vectors $e_u$ of elements in $\mathcal{F}(U)$. Then define $T: \mathcal{F}(U) \rightarrow U$ by $T(e_u) := v_u$ and extend linearly so that
\begin{equation}
	T(\sum \alpha_u e_u) = \sum \alpha_u v_u,
\end{equation}
where $e_u$ are the formal basis vector of the free vector space, $\{ v_u \}$ is an orthonormal set, and the $\alpha_u$ are scalar coefficients. Hence the map from a formal linear combination $\alpha u + \beta x \in \mathcal{F}(U_A)$ to a vector in the Hilbert space of the corresponding separated system $\psi_A = \alpha |u\rangle + \beta |x\rangle$ is explicitly
\begin{equation}
	T(\alpha e_u + \beta e_x) = \alpha T(e_u) + \beta T(e_x) = \alpha |u\rangle + \beta |x\rangle.
\end{equation}
Having now converted the formal linear combination in $\mathcal{F}(U_A)$ to a linear combination defined in terms of a vector basis, the vector $\alpha |u\rangle + \beta |x\rangle$ can be simplified if appropriate. Such simplification allows measurements to have a definite outcome. Explicit examples will be given below.

\section{Using the projection to separated subsystems}

We implement the projection as follows. Given $v_A, w_A \in \mathcal{H}_A$ and $v_B, w_B \in \mathcal{H}_B$, a maximally entangled state of the two subsystems, up to an overall phase and neglecting the normalization factor, is
\begin{equation}
	\Psi = v_A \otimes v_B + e^{i \phi} w_A \otimes w_B ,
\end{equation}
where $\phi$ defines the phase of the superposition. For the maximally-entangled states, we have $\phi = 2n\pi$ or $(2n+1)\pi$ (see Sec. 6). Recall from the last section that in the case of states of the composite system, for all the representatives of $\Psi$ in a coset of $\mathcal{F}$ the measurement outcomes are the same. In contrast, for measurements on the separated subsystems it turns out that the coset is subdivided into two classes of subsets (class 1 and class 2) that give \emph{different} measurement outcomes\cite{collapse}. 

When the superposition is an antisymmetric state ($\phi = (2n+1)\pi$), we have representatives:
\begin{align}
	\Psi  &= v_A \otimes v_B - (w_A \otimes w_B) \in \mathcal{H} \\
	&\rightarrow (v_A, v_B) + (w_A, -w_B) \quad (\text{class 1}) \in \mathcal{F} \nonumber \\
	\Psi  &= v_A \otimes v_B - (w_A \otimes w_B) \in \mathcal{H} \\
	&\rightarrow (v_A, v_B) + (-w_A, w_B) \quad (\text{class 2}) \in \mathcal{F} . \nonumber
\end{align}
The corresponding states projected into $\mathcal{H}_A$ and $\mathcal{H}_B$  that give definite measurement outcomes are:
\begin{eqnarray*}
	\psi_A (\text{class 1}) = v_A + w_A \\
	\psi_B (\text{class 1}) = v_B - w_B \\
	\psi_A (\text{class 2}) = v_A - w_A \\
	\psi_B (\text{class 2}) = v_B + w_B .
\end{eqnarray*}

When the superposition is a symmetric state ($\phi = 2n\pi$), we have:

\begin{align}
	\Psi  &= v_A \otimes v_B + w_A \otimes w_B \in \mathcal{H} \\
	&\rightarrow (v_A, v_B) + (w_A,  w_B) \quad (\text{class 1}) \in \mathcal{F} \nonumber \\
	\Psi  &= v_A \otimes v_B + w_A \otimes w_B \in \mathcal{H} \\
	&\rightarrow (v_A, v_B) + (-w_A, -w_B) \quad (\text{class 2}) \in \mathcal{F} . \nonumber
\end{align}
with
\begin{eqnarray*}
	\psi_A (\text{class 1}) = v_A + w_A \\
	\psi_B (\text{class 1}) = v_B + w_B \\
	\psi_A (\text{class 2}) = v_A - w_A \\
	\psi_B (\text{class 2}) = v_B - w_B .
\end{eqnarray*}

\section{Why there are two contextual phase classes}

A normalized state of a two-level system is written
\begin{equation}
	u = \alpha |0\rangle + \beta |1\rangle,
\end{equation}
where $\alpha, \beta \in \mathbb{C}$ and $|\alpha|^2 + |\beta|^2 = 1$. The vector $u$ can be identified with points lying on the circle $\mathbb{S}^3$ embedded in $\mathbb{C}^2 \cong \mathbb{R}^4$. 

Since states $u$ that differ by a global phase $e^{i\phi}$ are indistinguishable, we map all points in $\mathbb{S}^3$ that differ only by global phase to one point each on $\mathbb{S}^2$ by the Hopf fibration $\mathbb{S}^3/\mathbb{S}^1$, giving a surface of equivalence points known as the Bloch sphere. 

Consider now the superposition state
\begin{equation*}
	\psi_s = \frac{1}{\sqrt{2}} (u + w).
\end{equation*}
We can interpret the symmetric phase-locking of $u$ and $w$ to mean that their global phases are equal ($e^{i\phi_u} = e^{i\phi_w} $). Or, more generally, $|\phi_u - \phi_w| = 2n\pi$ with $n = 0, 1, 2,\dots$. Similarly an antisymmetric superposition,
\begin{equation*}
	\psi_a = \frac{1}{\sqrt{2}} (u - w),
\end{equation*}
can be interpreted as the superposition
\begin{equation}
	\psi_a = \frac{1}{\sqrt{2}} (e^{i\phi_u}u + e^{i\phi_w}w),
\end{equation}
where the global phases are antipodal. That is, $|\phi_u - \phi_w| = (2n+1)\pi$.

In each case, the superposition is produced by locking the global phases in- or out-of synchronization. This concept is easily generalized to arbitrary coefficients. 

Now consider the entangled states
\begin{equation*}
	\Psi_s = \frac{1}{\sqrt{2}} (u_A \otimes u_B + w_A \otimes w_B)
\end{equation*}
and 
\begin{equation*}
	\Psi_a = \frac{1}{\sqrt{2}} (u_A \otimes u_B - w_A \otimes w_B).
\end{equation*}
We can write either of these states in more general form as
\begin{equation}
	\Psi = \frac{1}{\sqrt{2}} (e^{i\xi_A + i\xi_B} u_A \otimes u_B + e^{i\phi_A + i\phi_B} w_A \otimes w_B),
\end{equation}
where we require for the symmetric linear combination $\Psi_s$ that
\begin{equation}
	|\xi_A + \xi_B  - \phi_A - \phi_B| = 2n\pi
\end{equation}
and for the antisymmetric state $\Psi_a$ that
\begin{equation}
	|\xi_A + \xi_B  - \phi_A - \phi_B| = (2n+1)\pi .
\end{equation}

There are infinitely many phase combinations for any state, but it turns out that we can divide these possibilities into two unique classes when there are two subsystems (A and B). The proof is as follows. Any maximally entangled state comprising two subsystems can be written up to a global phase for the state as 
\begin{equation*}
	\Psi = \frac{1}{\sqrt{2}} (u_A \otimes u_B + \lambda (w_A \otimes w_B)),
\end{equation*}
where $\lambda$ defines the phase of the superposition: $\lambda = e^{\phi_A + \phi_B  - \xi_A - \xi_B}$. Without loss of generality, we can rotate the arbitrary phases of the first term so that $\xi_A = \xi_B  = 0$, then focus on the phases of the second term. For a specific example consider a symmetric superposition with $\lambda = 1$. The same reasoning can then be extended to any other superposition. Since we require $\phi_A + \phi_B = 0$, we can choose phase pairs such as $\phi_A = \phi_B = 0$, $\phi_A = \frac{\pi}{4}$ and $\phi_B = -\frac{\pi}{4}$, and so on. In practice, the choice is irrelevant except when we perform measurements on the separated subsystems. In that case, any $\phi_A + \phi_B = 0$ with $\phi_A, \phi_B \in (-\frac{\pi}{2}, \frac{\pi}{2}]$ defines a particular measurement configuration. Notice that each $\phi_A$ (and $\phi_B$) has a corresponding antipodal vector in the opposite half of $\mathbb{S}^1$, that is for values $\phi_A, \phi_B \in (-\pi, -\frac{\pi}{2}] \cup (\frac{\pi}{2}, \pi]$. Since these vectors differ only by an overall phase, they cannot be discriminated by measurements---but they give complementary outcomes for measurements on the separated subsystems. Each vector has precisely one corresponding antipodal vector, and hence there are two equivalence classes of contextual phase.

The two phase classes are implied by conditions (4) and (5). Let's consider the case when $\lambda$ is real. Even multiples of $\pi$ mean that the equivalence classes of phases are represented by $e^{\phi_A}$ and $e^{\phi_B}$ are both equal to 1 or are both $-1$. Odd multiples of $\pi$ mean that the equivalence classes of phases are represented by one phase factor having the value 1 and the other $-1$. Complex phases satisfy similar conditions, as is easily seen. These two unique phase combinations derive from how bilinear maps become linear maps in a tensor product space. They define the two contextual phase classes. 

\section{Example}
Let's review a specific example. As is well known, entangled state wavefunctions can be written in different-looking, but equivalent forms\cite{EkertKnight}. These different forms are related by unitary transformations on the composite state $\Psi$. Note that usually the $x$-basis form of $\Phi_-$ is labeled $\Psi_+$ and, conversely, the $x$-basis form of $\Psi_+$ is labeled $\Phi_-$. Here we do not change the label. Instead we keep the $z$-basis notation regardless of the basis in order to emphasize that these wavefunctions are identical.

We observe that we can choose a representation of $\Psi$ compatible with a chosen measurement basis so that measurement outcomes in that basis are well-defined. Consider $\Psi_+$. We can present it equivalently in the following form:
\begin{equation*}
	\Psi_+ = \frac{1}{\sqrt{2}} \Big[ |0'\rangle_A \otimes |0'\rangle_B - |1'\rangle_A \otimes |1'\rangle_B  \Big] ,
\end{equation*}
where $|0'\rangle \equiv \frac{1}{\sqrt{2}}(|0\rangle + |1\rangle)$, $|1'\rangle \equiv \frac{1}{\sqrt{2}}(|0\rangle - |1\rangle)$ and below we will also refer to $|-0'\rangle \equiv \frac{1}{\sqrt{2}}(-|0\rangle - |1\rangle)$, and $|-1'\rangle \equiv \frac{1}{\sqrt{2}}(-|0\rangle + |1\rangle)$.  

We can re-write this informally as a pre-image, with a class 1 contextual phase choice, to foreshadow the class 1 representation in $\mathcal{F}$, but retain the familiar tensor product notation:
\begin{equation}
	\Psi_+ = \frac{1}{\sqrt{2}} \Big[ |0'\rangle_A \otimes |0'\rangle_B + |1'\rangle_A \otimes |-1'\rangle_B  \Big] .
\end{equation}
Then it is apparent that the canonical map on the corresponding formal linear combination in $\mathcal{F}(U_A \times V_B)$ gives the formal linear combination in $\mathcal{F}(U_A)$:
\begin{equation}
	\psi_A = \frac{1}{2}  ( e_0 + e_1 + e_0 - e_1  ),
\end{equation}
which maps to the vector in $\mathcal{H}_A$
\begin{equation*}
	\psi_A = \frac{1}{2} ( |0\rangle + |1\rangle + |0\rangle - |1\rangle  )  .
\end{equation*}
Notice that the vector $\psi_A$ simplifies when a measurement is made on the separated subsystem to give a definite outcome based on the expectation value for the operator acting on $|0\rangle$. That is, for some operator $\mathcal{O}$,  $\langle \psi_A | \mathcal{O} | \psi_A \rangle = \langle 0 | \mathcal{O} | 0 \rangle$.

However, the vector does not simplify or collapse simply because the subsystems are separated. If that happened, then the entanglement would be destroyed and we are left with a classical state. We know this must be the case because entanglement between widely separated systems, such as photons, is maintained. The simplification, or collapse, of this projected state vector only happens when a measurement is performed. The mechanism by which this happens is an interference effect in the measuring device, whereby one outcome (in this example $|0\rangle$) is generated by constructive interference and the other outcome ($|1\rangle$) is suppressed by destructive interference. 

Similarly for a measurement on subsystem B, in $\mathcal{H}_B$, where the canonical map on the corresponding formal linear combination in $\mathcal{F}(U_A \times V_B)$ gives the formal linear combination in $\mathcal{F}(V_B)$:
\begin{equation}
	\psi_B = \frac{1}{2}  ( e_0 + e_1 - e_0 + e_1  ),
\end{equation}
which maps to the vector in $\mathcal{H}_B$
\begin{equation*}
	\psi_B = \frac{1}{2}   ( |0\rangle + |1\rangle - |0\rangle + |1\rangle )  .
\end{equation*}

So that for some operator $\mathcal{O}$,  $\langle \psi_B | \mathcal{O} | \psi_B \rangle = \langle 1 | \mathcal{O} | 1 \rangle$. Working through the case of $\Psi_+(\text{class 2})$, we obtain the opposite outcomes, that is, $\langle \psi_A | \mathcal{O} | \psi_A \rangle = \langle 1 | \mathcal{O} | 1 \rangle$  and $\langle \psi_B | \mathcal{O} | \psi_B \rangle = \langle 0 | \mathcal{O} | 0 \rangle$. Hence we obtain the same expectation value for measurements on subsystem A as predicted by Eq. 5:
\begin{align}
	E_A &=  \frac{1}{2} \langle \psi_A^{\text{class 1}} | T_A | \psi_A^{\text{class 1}} \rangle + \frac{1}{2} \langle \psi_A^{\text{class 2}} | T_A | \psi_A^{\text{class 2}} \rangle \\
	&= \frac{1}{2} \langle 0 | T_A | 0 \rangle + \frac{1}{2} \langle 1 | T_A | 1 \rangle. \nonumber
\end{align}

\section{Contextual phase is associated with measurement basis}

Any Bell state can be written in the $z$-basis (indicated by the superscript) in two equivalent representative ways that differ only in the contextual phase of the second term. the contextual phase class 1 or 2 is indicated in the superscript. The symmetric states are:
\begin{eqnarray*}
	\Phi^{(1,z)}_+ = \frac{1}{\sqrt{2}} \Big[ |0\rangle_A \otimes |0\rangle_B + |1\rangle_A \otimes |1\rangle_B  \Big] \\
	\Phi^{(2,z)}_+ = \frac{1}{\sqrt{2}} \Big[ |0\rangle_A \otimes |0\rangle_B + |-1\rangle_A \otimes |-1\rangle_B  \Big] \\
	\Psi^{(1,z)}_+ = \frac{1}{\sqrt{2}} \Big[ |0\rangle_A \otimes |1\rangle_B + |1\rangle_A \otimes |0\rangle_B  \Big] \\
	\Psi^{(2,z)}_+ = \frac{1}{\sqrt{2}} \Big[ |0\rangle_A \otimes |1\rangle_B + |-1\rangle_A \otimes |-0\rangle_B  \Big]
\end{eqnarray*}
and the antisymmetric states are:
\begin{eqnarray*}
	\Phi^{(1,z)}_- = \frac{1}{\sqrt{2}} \Big[ |0\rangle_A \otimes |0\rangle_B + |-1\rangle_A \otimes |1\rangle_B  \Big] \\
	\Phi^{(2,z)}_- = \frac{1}{\sqrt{2}} \Big[ |0\rangle_A \otimes |0\rangle_B + |1\rangle_A \otimes |-1\rangle_B  \Big] \\
	\Psi^{(1,z)}_- = \frac{1}{\sqrt{2}} \Big[ |0\rangle_A \otimes |1\rangle_B + |-1\rangle_A \otimes |0\rangle_B  \Big] \\
	\Psi^{(2,z)}_- = \frac{1}{\sqrt{2}} \Big[ |0\rangle_A \otimes |1\rangle_B + |1\rangle_A \otimes |-0\rangle_B  \Big]
\end{eqnarray*}

We observe that there is no continuous map moving solely on the Hopf fibers $\mathbb{S}^1_A \times \mathbb{S}^1_B$ (i.e. keeping the basis for the state fixed) that converts a class 1 phase configuration to a class 2 phase configuration because conditions (4) or (5) would be broken. However, synchronous, continuous rotation of each of the vectors on $\mathbb{S}^3_A \times \mathbb{S}^3_B$ changes the basis cyclically \emph{and} changes the phase class cyclically, without changing the state. 

To illustrate this result, consider vectors with real coefficients and the usual basis vectors $|0\rangle = 
\begin{pmatrix}
	1 \\
	0  
\end{pmatrix} $
and $|1\rangle = 
\begin{pmatrix}
	0 \\
	1  
\end{pmatrix} $. We can rotate these basis vectors by applying rotations using the usual rotation matrix:
\begin{equation}
	R_{\theta} = 
	\begin{pmatrix}
		\cos \theta & -\sin \theta \\
		\sin \theta & \cos \theta 
	\end{pmatrix}.
\end{equation}
We thus define the operator $U = R_{\zeta} \otimes R_{\theta}$ such that 
\begin{equation}
	U(u \otimes w) = (R_{\zeta} u) \otimes (R_{\theta} w).
\end{equation}

Now consider joint measurements on various common bases. Set $\zeta = \theta = \frac{\pi}{4}$. Applying the same rotation to both subsystems leaves the state unchanged. Only the basis in which  we determine or measure the wavefunction changes.  Because the coefficients are real, we  can visualize them  on $\mathbb{S}^1_A \times \mathbb{S}^1_B$. A sequence of $\frac{\pi}{4}$ rotations generates a cycle of states and contextual phase classes. For instance, consider $\Phi_+$. Start by writing down the state in the $z$-basis and then apply $U$ sequentially to obtain

\begin{tikzpicture}[node distance = 3cm, auto, >= latex]
	\node[font=\LARGE] (A) at (90 : 3cm) {$\mathcolor{red}{\Phi^{(1,z)}_+}$};
	\node[font=\LARGE] (B) at (45 : 3cm) {$\mathcolor{blue}{\Phi^{(2,x)}_+}$};
	\node[font=\LARGE] (C) at (0 : 3cm) {$\bar{\Phi}^{(2,z)}_+$};
	\node[font=\LARGE] (D) at (315 : 3cm) {$\bar{\Phi}^{(1,x)}_+$};
	\node[font=\LARGE] (E) at (270 : 3cm) {$\bar{\Phi}^{(1,z)}_+$};
	\node[font=\LARGE] (F) at (225 : 3cm) {$\bar{\Phi}^{(2,x)}_+$};
	\node[font=\LARGE] (G) at (180 : 3cm) {$\mathcolor{red}{\Phi^{(2,z)}_+}$};
	\node[font=\LARGE] (H) at (135 : 3cm) {$\mathcolor{blue}{\Phi^{(1,x)}_+}$};
	\draw [->, thick] (A) to [bend left = 20] (B);
	\draw [->, thick] (B) to [bend left = 20] (C);
	\draw [->, thick] (C) to [bend left = 20] (D);
	\draw [->, thick] (D) to [bend left = 20] (E);
	\draw [->, thick] (E) to [bend left = 20] (F);
	\draw [->, thick] (F) to [bend left = 20] (G);
	\draw [->, thick] (G) to [bend left = 20] (H);
	\draw [->, thick] (H) to [bend left = 20] (A);
\end{tikzpicture}

Where:
\begin{align*}
	\Phi^{(1,z)}_+ &= \frac{1}{\sqrt{2}} \Big[ 
	\begin{pmatrix}
		0 \\
		1  
	\end{pmatrix}  
	\begin{pmatrix}
		0 \\
		1  
	\end{pmatrix}
	+
	\begin{pmatrix}
		1 \\
		0  
	\end{pmatrix}  
	\begin{pmatrix}
		1 \\
		0  
	\end{pmatrix}
	\Big] \\
	\Phi^{(2,x)}_+ &= \frac{1}{2\sqrt{2}} \Big[ 
	\begin{pmatrix}
		-1 \\
		1  
	\end{pmatrix}  
	\begin{pmatrix}
		-1 \\
		1  
	\end{pmatrix}
	+
	\begin{pmatrix}
		1 \\
		1  
	\end{pmatrix}  
	\begin{pmatrix}
		1 \\
		1  
	\end{pmatrix}
	\Big] \\
	\bar{\Phi}^{(2,z)}_+ &= \frac{1}{\sqrt{2}} \Big[ 
	\begin{pmatrix}
		-1 \\
		0  
	\end{pmatrix}  
	\begin{pmatrix}
		-1 \\
		0  
	\end{pmatrix}
	+
	\begin{pmatrix}
		0 \\
		1  
	\end{pmatrix}  
	\begin{pmatrix}
		0 \\
		1  
	\end{pmatrix}
	\Big] \\
	\bar{\Phi}^{(1,x)}_+ &= \frac{1}{2\sqrt{2}} \Big[ 
	\begin{pmatrix}
		-1 \\
		-1  
	\end{pmatrix}  
	\begin{pmatrix}
		-1 \\
		-1  
	\end{pmatrix}
	+
	\begin{pmatrix}
		-1 \\
		1  
	\end{pmatrix}  
	\begin{pmatrix}
		-1 \\
		1  
	\end{pmatrix}
	\Big] \\
	\bar{\Phi}^{(1,z)}_+ &= \frac{1}{\sqrt{2}} \Big[ 
	\begin{pmatrix}
		0 \\
		-1  
	\end{pmatrix}  
	\begin{pmatrix}
		0 \\
		-1  
	\end{pmatrix}
	+
	\begin{pmatrix}
		-1 \\
		0  
	\end{pmatrix}  
	\begin{pmatrix}
		-1 \\
		0  
	\end{pmatrix}
	\Big] \\
	\bar{\Phi}^{(2,x)}_+ &= \frac{1}{2\sqrt{2}} \Big[ 
	\begin{pmatrix}
		1 \\
		-1  
	\end{pmatrix}  
	\begin{pmatrix}
		1 \\
		-1  
	\end{pmatrix}
	+
	\begin{pmatrix}
		-1 \\
		-1  
	\end{pmatrix}  
	\begin{pmatrix}
		-1 \\
		-1  
	\end{pmatrix}
	\Big] \\
	\Phi^{(2,z)}_+ &= \frac{1}{\sqrt{2}} \Big[ 
	\begin{pmatrix}
		1 \\
		0 
	\end{pmatrix}  
	\begin{pmatrix}
		1 \\
		0 
	\end{pmatrix}
	+
	\begin{pmatrix}
		0 \\
		-1  
	\end{pmatrix}  
	\begin{pmatrix}
		0 \\
		-1  
	\end{pmatrix}
	\Big] \\
	\Phi^{(1,x)}_+ &= \frac{1}{2\sqrt{2}} \Big[ 
	\begin{pmatrix}
		1 \\
		1  
	\end{pmatrix}  
	\begin{pmatrix}
		1 \\
		1  
	\end{pmatrix}
	+
	\begin{pmatrix}
		1 \\
		-1  
	\end{pmatrix}  
	\begin{pmatrix}
		1 \\
		-1  
	\end{pmatrix}
	\Big] .
\end{align*}

The cycle of basis transformations of a wavefunction, such as illustrated above, shows the full structure of an entangled state. Not only does the wavefunction encode various basis-dependent joint measurement outcomes, it encodes also the correlations and apparent randomness characteristic of these measurements. Randomness comes from the fact that we do not know the contextual phase class for a reference measurement setting (i.e. a measurement that is not made, but could be in principle).  Thus the measurement outcomes for measurements in any basis appear random when viewed through the lens of classical measurement devices. 

The set of possible measurement outcomes are not collected in the wavefunction such that they can each be recorded like a list of classical properties---as is well known. Instead, any one of these outcomes, subject to the ambiguity of class 1 and 2 contextual phases, can be elucidated by suitable configuration of the measurement apparatus. Observed joint correlations depend on how we `view' the wavefunction projected into each subsystem via the pair of detector settings. 

Notice that conditions (4) and (5) show that we are free to measure each subsystem completely independently for a pair of basis choices. Consider Eq. (7) extended linearly---choosing any pair of $\zeta$ and $\theta$. The difference between $\zeta$ and $\theta$ is guaranteed to be compensated in conditions (4) or (5).  Correlations in the measurement outcomes do not depend on the result of the first measurement; which is well-known to be an unresolved issue for proponents of the collapse model. Instead, these measurement results are there all along in the wavefunction. How they are revealed to an observer depends only on the measurement settings. We might think of the wavefunction as a very special kind of intrinsic ensemble, which ultimately derives from its wavelike foundation.

The \emph{apparent} randomness of the first measurement outcome is confusing because because it obscures the fact that the second outcome is intrinsically associated to the first outcome according to the contextual phase class. For example, given a specific wavefunction, such as $\Psi^{(1,x)}_-$, the two measurement outcomes are completely predetermined for any pair of measurement settings. The appearance of random outcomes comes from the observer not being able to set or even know the contextual phase class.

\section{Discussion}

The results of this work imply that there is an additional phase structure in entangled states, relating to the phase of the superposition between simple tensors, that is irrelevant (indeed, invisible) for any map on $\Psi$ itself, but that may be revealed by measurements on the separated subsystems. This additional structure is evident in the free vector space that produces the tensor product space by the quotient. The interpretation suggested by this observation is that an ensemble of \emph{identical} wavefunctions $\Psi$ can yield an ensemble of measurement pairs $\langle \psi_A | \mathcal{O} | \psi_A \rangle$ and $\langle \psi_B | \mathcal{O} | \psi_B \rangle$. The ensemble is divided into two classes, such that measurement outcomes, in the case of $z$-basis measurements on $\Psi_+$ or $\Psi_-$, are 0 and 1 for subsystems A and B respectively in class 1 cases, but are 1 and 0 for class 2 cases. From this perspective, there is no need to invoke random collapse of the superposition to ensure the measurement outcomes are definite. 

Replacing the random collapse hypothesis with a specialized statistical model for measurement outcomes (but \emph{not} for the $\Psi$) does not mean that the correlations in measurement outcomes are locally-encoded correlations. The outcomes depend on the basis chosen for the first measurement. This is essentially the same concept as `Mermin's square'\cite{Mermin1993, Peres, Svozil2021}. For example, if we choose to measure subsystem A in the $z-$basis, then the entire $\Psi$ must be represented in a form that allows us to obtain a definite outcome. Then a measurement on subsystem B \emph{in this same basis} has a predictable outcome. Instead, if we attempt to measure B in the $x-$basis, we will find a random outcome. The measurement basis plays a central role, and therefore the nonlocality associated with these measurements is that the measurement basis chosen for the first measurement on a subsystem controls the form of $\Psi$, no matter how far apart these subsystems are. 

The nonlocal effect is not an interaction, which would be difficult to explain, as has been widely discussed. Instead, the insight gained from our model is that these correlated outcomes are subtly pre-determined in the wave-like structure of a quantum state. They are intrinsically related to the basis representation of the state, as described in Sec. 8. 

In sum, the usual viewpoint for the source of nonlocality is that collapse of the superposition of $\psi_A$ somehow controls the collapse of a superposition of $\psi_B$ for a second subsystem, remote from the first. This concerted collapse explains the correlations between measurement outcomes. The present model proposes that correlated measurement outcomes are embedded statistically into the $\Psi$, where this ensemble is only revealed by measurements on the separated subsystems---it is invisible to any measurement on $\Psi$. In the latter model, the curious nonlocal effect can be explained by how the superposition principle manifests in the projections of states of the separated subsystems . 

\section{Conclusions}

Measurements on entangled states were studied. In particular, we examined measurements on separated subsystems associated with an entangled state.  It was noted that measurements on states of the composite system only give one measurement outcome. Whereas, measurements on the separated subsystems can give a pair of outcomes. The crucial idea is that physical separation of the subsystems enables the state of a composite system to `project' into the Hilbert spaces local to the subsystems. Based on the definition of the tensor product, it was shown how to obtain such a `projection' by moving from the tensor product representation of the vector to the corresponding coset in the associated free vector space. Theorem 1 gives a plausible basis for clarifying the nature of nonlocal correlations in joint measurements on entangled states. By eliminating random collapse, the results suggest that nonlocality arises because of the way the superposition principle applies to projections of composite states with complementary contextual phase class. The measurement basis probes different possible superpositions associated to the separated subsystems. 

The work clarifies key issues in quantum foundations including: (i) The basis for how measurements on separated subsystems of a composite state give an outcome for each subsystem, whereas similar measurements on the composite state can only give one outcome; (ii) The physical basis for the collapse postulate; (iii) How measurements on separated subsystems give more detail about quantum correlations than measurements on the composite state; (iv) Why nonlocal correlations arise without any kind of interaction between the subsystems.


\medskip
\textbf{Acknowledgements} \par 
This research was funded by the Division of Chemical Sciences, Geosciences and Biosciences, Office of Basic Energy Sciences,of the US Department of Energy through grant no. DE-SC0015429.

\medskip

%
\bibliographystyle{MSP}
\bibliography{Scholes_bib_July2026}


\end{document}